\documentclass[seceq,jjapprn]{jjap}
\usepackage[dvips]{epsfig}
\newcommand{\error}[2]{\mbox{$#1 \pm #2$}}

\title{Proton Irradiation Experiment for the X-ray Charge-Coupled
Devices of the Monitor of All-sky X-ray Image mission onboard the
International Space Station: I. Experimental Setup and Measurement of
the Charge Transfer Inefficiency}

\author{Emi {\sc Miyata}, Tomoyuki {\sc Kamazuka}, Hirohiko
{\sc Kouno}, Mitsunori {\sc Fukuda}$^{1}$, Mototsugu {\sc
Mihara}$^{1}$, Kensaku {\sc Matsuta}$^{1}$, Hiroshi {\sc Tsunemi},
Kanenobu {\sc Tanaka}$^{1}$, Tadanori {\sc Minamisono}$^{1}$, Hiroshi
{\sc Tomida}$^{2}$ and Kazuhisa {\sc Miyaguchi}$^{3}$}

\inst{
Department of Earth \& Space Science, 
Graduate School of Science, Osaka University, \\
1-1 Machikaneyama-cho, Toyonaka, Osaka 560-0043, Japan\\
$^1$ Department of Physics, 
Graduate School of Science, Osaka University, \\
1-1 Machikaneyama-cho, Toyonaka, Osaka 560-0043, Japan\\
$^2$ National Space Development Agency of Japan (NASDA)\\
Tsukuba Space Center, 
2-1-1 Sengen Tsukuba Ibaragi 305-8505, Japan\\
$^3$ Solid State Division, Hamamatsu Photonics K.K., \\
1126-1 Ichino-cho, Hamamatsu City 435-8558, Japan
}

\abst{We have investigated the radiation damage effects on a CCD to be
employed in the Japanese X-ray astronomy mission including the Monitor
of All-sky X-ray Image (MAXI) onboard the International Space Station
(ISS).  Since low energy protons release their energy mainly at the
charge transfer channel, resulting a decrease of the charge transfer
efficiency, we thus focused on the low energy protons in our
experiments.  A 171\,keV to 3.91\,MeV proton beam was irradiated to a
given device. We measured the degradation of the charge transfer
inefficiency (CTI) as a function of incremental fluence. A 292\,keV
proton beam degraded the CTI most seriously.  Taking into account the
proton energy dependence of the CTI, we confirmed that the transfer
channel has the lowest radiation tolerance.  We have also developed the
different device architectures to reduce the radiation damage in
orbit. Among them, the ``notch'' CCD, in which the buried channel
implant concentration is increased, resulting in a deeper potential
well than outside, has three times higher radiation tolerance than that
of the normal CCD. We then estimated the charge transfer inefficiency of
the CCD in the orbit of ISS, considering the proton energy spectrum.
The CTI value is estimated to be $1.1 \times 10^{-5}$ per each transfer
after two years of mission life in the worse case analysis if the
highest radiation-tolerant device is employed. This value is well within
the acceptable limit and we have confirmed the high radiation-tolerance
of CCDs for the MAXI mission.}

\kword{charge-coupled device, radiation damage, displacement,
international space station, radiation belt}

\begin{document} 

 \maketitle 
 \sloppy

 \section{Introduction}

 Charge-coupled devices (CCDs) have emerged as the preferred detectors
 on all new X-ray astronomy mission in recent years. This is because
 they possess a high spatial resolution as well as a moderate energy
 resolution, simultaneously~\cite{tanaka}. The dead layer above CCD must
 be thin enough to attain a high quantum efficiency at soft X-ray
 regions. As the result, devices cannot be protected against the high
 energy particles in space in the incident direction of X-rays.

 Soon after the launch of the X-ray astronomy satellite, {\it Chandra},
 all of the front-illuminated CCD of the X-ray CCD camera (ACIS) have
 suffered some damage caused by the charge transfer inefficiency
 (CTI)~\cite{acis_damage}.  The CTI is defined as an average fraction of
 charge packet lost at each transfer. Similar type of devices to ACIS
 CCDs have been tested by the high energy protons (40\,MeV and 10\,MeV)
 but not by the low energy protons before launch. The low energy protons
 having energy of $\sim\,$150\,keV release major part of its energy at
 the transfer channel of the ACIS CCDs, which is located roughly
 1\,$\mu$m below the electrodes. They cause the displacement damages in
 Si, leading to the formation of trapping sites for the charge
 packet. Since the flux of low energy protons at the orbit of {\it
 Chandra} is much higher than that at the low earth orbit such as
 ASCA~\cite{acis_proton} and low energy protons reflecting through the
 X-ray mirror assembly (HRMA) can reach the focal
 plane~\cite{acis_reflect}, a significant degradation of the CTI has
 occurred.

 The Monitor of All-sky X-ray Image (MAXI) has been selected as an early
 payload of the JEM (Japanese Experimental Module; {\it KIBO}) Exposed
 Facility on the International Space Station (ISS)~\cite{maxi}. MAXI has
 slit scanning cameras which consist of two kinds of X-ray detectors;
 one-dimensional position sensitive proportional counters with total
 area of $\sim$\,5000\,cm$^2$ named GSC and the X-ray CCD camera with
 total area of $\sim$\,200\,cm$^2$ named SSC.  SSC carries 32 CCDs which
 are three-side buttable with full-frame transfer and have
 1024$\times$1024 pixels of $24\,\mu {\rm m}\times24\,\mu {\rm m}$ size
 with two phase gate structures. The CCD chips are fabricated by the
 Hamamatsu Photonics K. K. (HPK). In order to perform useful X-ray
 spectroscopy over the whole device, the CTI must be less than roughly
 $2\times 10^{-5}$\, per transfer where the shift of the peak energy is
 similar to that of the Fano-limited noise of 120\,eV at 5.9\,keV.

 Previous studies of the radiation hardness for HPK CCDs were also
 focused on high energy protons above 1\,MeV~\cite{tomida} and no data
 are available for low energy protons. We thus performed the irradiation
 test focusing on the low energy protons. In this paper, we describe the
 device architecture, irradiation experiment, and the measurement of the
 CTI at $-100\,^\circ$C.

 \section{Experiments and Results}

 \subsection{Architecture of CCD}\label{sec:wafer}

 We employed CCDs packaged in a half-inch size which is different from
 that of the MAXI CCD whereas the device was fabricated from the same
 type of wafer of the MAXI CCD.  The devices irradiated possess
 512$\times 512$ pixels of which size is 24 $\mu$m
 square. Figure~\ref{fig:ccd_architecture}(a) shows the schematic view
 of the CCD. The cross section of a CCD pixel in the horizontal and the
 vertical direction is shown in Fig~\ref{fig:ccd_architecture}(b)
 (we call this type of device as normal CCD which has
 no countermeasure for radiation hardness).  Since the CCD employed is a
 buried-channel type, there is a thin ($\simeq 1\,\mu$m) doped $n$-type
 layer just below the oxide layer.  There are four electrodes per pixel
 in our case and they slightly overlap each other as shown in the cross
 section of Fig~\ref{fig:ccd_architecture}(b) ({\it right}). There is
 another gate structure, a ``channel stop'', which forms the pixel
 boundary along the vertical transfer direction shown in
 Fig~\ref{fig:ccd_architecture}(b) ({\it left}).

 We need to develop a highly radiation-tolerant device to be employed in
 space for a long time. One of the design to improve the radiation
 tolerance is a ``notch'' structure~\cite{janesick}. The notch CCD
 incorporated a narrow, $\sim \,3\, \mu$m, strip both for vertical and
 horizontal registers in which the buried channel implant concentration
 was increased, resulting in a deeper potential well than outside the
 notch structure as shown in bottom of
 Fig~\ref{fig:ccd_architecture}(c).  Small charge packet would be
 confined within the notch structure to reduce the probability to
 encounter with vacant traps by a factor of $\sim$8. We therefore expect
 that the CTI would be improved in the device having the notch structure
 as CRAF/Cassini CCDs~\cite{janesick} and EEV CCDs~\cite{holland}.  We
 should note that all devices used in section~\ref{sec:cti} and
 \ref{sec:bias} possesses the notch structure.

 Another technique employed is to replace some amount of the Si oxide
 with the nitride oxide as shown in Fig~\ref{fig:ccd_architecture}(d)
 (nitride CCD, hereafter). The thickness of the oxide layer and the
 nitride layer of the nitride CCD we fabricated is similar to that of
 the oxide layer of the normal device. There is no differences below and
 above the nitride and oxide layer between the nitride CCD and the
 normal CCD. The nitridation of the oxide layer enables us to reduce the
 ionization damage, resulting the effect of the flat-band shift to be
 smaller than those of devices having only oxides~\cite{nitride1,
 nitride2}.  This technique would reduce the dark current for damaged
 devices.  It might not be efficient to improve the CTI since the
 nitride CCD possesses the similar structure of the depletion layer to
 the normal CCD.  However, the degradation of the CTI must be similar
 level for normal devices.

 We have developed CCDs from various Si wafers.  Details about
 newly-developed devices have been described in Miyata {\it et
 al.}~\cite{ssc_em}.  Devices fabricating from epitaxial wafer and from
 bulk wafer were tested.  We have decided to employ devices from
 epitaxial wafer in order to achieve the high energy resolution, high
 quantum efficiency for hard X-rays, and also low dark current.  Among
 them, we employed the epitaxial-2 (hereafter referred as epi-2) and
 epitaxial-3 (referred as epi-3) devices for comparison. The resistivity
 of epi-3 wafer is roughly an order of magnitude higher than that of
 epi-2 wafer.

 We then fabricated following four types devices to compare their
 radiation tolerance; epi-3 with and without notch, epi-2 with notch,
 and epi-2 without notch but having the nitride oxide.

 \subsection{Experimental setup and beam calibration}

 A 570\,keV to 4.0\,MeV proton beam, shown in
 Table~\ref{table:proton_energy}, was provided by the Van de Graaff
 accelerator at the Laboratory for Nuclear Studies, Osaka University.
 We employed an Al degrader with a thickness of 5\,$\mu$m to reduce the
 energy down to 171 keV (Table~\ref{table:proton_energy}). Pulsed beams
 were used to control the irradiation fluence. The proton beam was
 over-defocussed by quadrupole magnet to obtain a weak and uniform beam
 at the CCD.

 Figure~\ref{fig:degrador-ccd} shows the experimental setup of the CCD
 chamber.  We employed four diaphragm plates made of Al between the
 degrader and the CCD in order to reduce scattered protons or secondary
 electrons generated by the inside wall of the duct. Inside the CCD
 chamber, the collimator made of Al with a thickness of 3\,mm having a
 hole of 21\,mm\,$\phi$ was set both to collimate the proton beam and to
 monitor the intensity of the incident proton beam to the CCD. Two
 $^{55}$Fe sources were located behind the collimator. It enabled us to
 investigate the {\it in-situ} performance of the CCD.

 Roughly a half area of each CCD was shielded against protons to provide
 the non-irradiated data to compare with those from the irradiated
 region. Since the amount of scattering of protons is not small, the
 vertical boundary between the shielded and the irradiated region was
 not so clear.  It makes the calculation of the CTI in the horizontal
 transfer direction to be much uncertain.  We thus focused on the
 vertical CTI only.

 We drove CCDs with our newly-developed system, {\sl E-NA}
 system~\cite{e-na}. The CCD analog data were processed by an
 integration-type circuit~\cite{ssc_em} and digital data were acquired
 with a VME system. The readout noise of our system including the device
 is less than 10\,e$^-$ rms under the charge transfer rate of 125\,kHz.

 We calibrated the beam current with replacing the CCD by a Faraday cup
 made of Al shown in Fig~\ref{fig:degrador-ccd}. We measured the current
 at the Faraday cup and the collimator with pico-ammeter,
 simultaneously. We thus obtained the ratio of each current at given
 energy. We found that the ratio depended on the beam energy possibly
 because the probability of secondary emission depends on the beam
 energy when protons hit materials inside the chamber.  The accuracy of
 the beam intensity estimated from the collimator current could be
 achieved to be 5\,\% taking into account their energy dependence.

 The spatial uniformity of the beam intensity was measured with 650\,keV
 protons. The proton energy at downstream of the degrader was 292\,keV
 (detail will be described in the next section).  A 292\,keV proton
 generate electron-hole pair along its track inside the CCD. Electrons
 generated only at the depletion layer can be collected. After passing
 through the covering layer, electrode, and the oxide layer shown in
 Fig~\ref{fig:ccd_architecture}(b), the proton energy was reduced to be
 $\sim 6$\,keV which generate charge cloud consisting of $\sim 1.6\times
 10^3$ electrons, which is much lower than the capacity of a pixel.  We,
 therefore, irradiated very week 650\,keV proton beam onto the CCD so
 that pile-up effect is negligibly small. Protons detected by CCDs
 formed various types of events similar to X-rays: single-pixel event,
 two-pixel split event, and more extended event. We thus measured the
 number of events generated by protons in 24\,mm squares and found the
 spatial uniformity to be better than 10\,\%.

 \subsection{Performance degradation of mono-energetic protons}\label{sec:cti}

 We performed the incremental irradiations at given energy on a given
 device. All devices were in operation (biased) during this experiment
 and fabricated from the same wafer (epitaxial-3 wafer).
 Table~\ref{table:proton_energy} shows the energy of proton
 irradiated. The center energy and width of proton spectrum at
 downstream of the degrader were calculated with the {\sl
 Geant4}~\cite{geant4}. We employed the low energy extension with {\sl
 G4EMLOW0.3} data in {\sl Geant4} in order to simulate the physical
 process of low energy less than 2~MeV~\cite{geant4_le}. In the
 subsequent section, we only referred the proton energy to that at
 downstream of the degrader.

 Figure~\ref{fig:specs}\,(a) shows the spectrum of $^{55}$Fe extracted
 from single-pixel events before proton irradiation. The energy
 resolution of Mn K$\alpha$ has a full-width at a half maximum (FWHM) of
 146 eV. After an irradiation of 292\,keV proton with a fluence of
 $1.04\times 10^7$\,cm$^{-2}$, the degradation of the detector
 performance was significant and the energy resolution became 294 eV
 (Fig~\ref{fig:specs}\,(b)). The peak positions of Mn K$\alpha$
 and K$\beta$ were shifted, suggesting the incomplete collection of the
 charge packet. We then incremented fluence up to $1.11 \times
 10^8$\,cm$^{-2}$ and resultant spectrum is shown in
 Fig~\ref{fig:specs}\,(c). Mn K$\alpha$ and K$\beta$ X-rays could not be
 resolved and the energy resolution was degraded to 614 eV. The device
 irradiated with 292\,keV protons suffered the most serious damage on
 the energy resolution compared with those irradiated by protons of
 other energies.

 Figure~\ref{fig:readout_noise} shows the readout noise as a function of
 proton fluence for 292\,keV and 3.91\,MeV protons.  Since the readout
 noise was evaluated from the histogram of the horizontal over-clocked
 area, it included the noise of CCD as well as that of the electronics.
 Before the proton irradiation, the readout noise of both CCDs were
 $\sim$\,7.8\,e$^-$. Therefore, there was not an influence of proton
 irradiations on the readout noise.  In this way, we confirmed no
 degradation of the readout noise for irradiations with protons of any
 energies employed. The degradation of energy resolution shown in
 Fig~\ref{fig:specs} was not caused by the degradation of the readout
 noise.

 Figure~\ref{fig:pha_rawy} shows the pulse heights of $^{55}$Fe events
 as a function of the number of transfer. Each dot in these figures
 corresponds to an individual X-ray event. The histogram shown in
 Fig~\ref{fig:specs} can be obtained if one makes a projection of these
 plots to Y-axis (pulse height axis).  Before irradiation, there are two
 horizontal lines clearly seen, corresponding to Mn K$\alpha$ at $\sim$
 710 channel and K$\beta$ at $\sim$ 790 channel
 (Fig~\ref{fig:pha_rawy}\,(a)).  After irradiating with 292\,keV protons
 with fluence of $1.04\times 10^7$\,cm$^{-2}$, the pulse height of X-ray
 events decreases with increasing transfer number, suggesting the loss
 of the charge packet during the transfer
 (Fig~\ref{fig:pha_rawy}\,(b)). We should note that the widths of two
 lines were broadened as the transfer number became larger.
 Figure~\ref{fig:pha_rawy}\,(c) shows the same plot after irradiating
 with protons of $1.11 \times 10^8$\,cm$^{-2}$. The significant loss of
 charge packet is found and the pulse height at the transfer number of
 500 is less than half of that before irradiation. The pulse height at
 the transfer number of zero is still less than that before irradiation,
 suggesting the loss of charge packet in the serial register of the
 device.

 In order to characterize the loss of charge packet, we calculated the
 values of the CTI for all proton energies based on
 Fig~\ref{fig:pha_rawy}.  Figure~\ref{fig:cti} shows the CTI as a
 function of proton fluence for various proton energies. Protons having
 energy of 150\,keV have seriously degraded the detector performance in
 the case of ACIS. On the other hand, 171\,keV protons affected the CTI
 for HPK CCDs less effectively. Instead, HPK CCDs suffered serious
 damage by protons with higher energies of 292 and 391\,keV. The
 degradation of the CTI caused by proton energies above 500\,keV is
 again less than those of protons of 292 and 391\,keV.

 Since values of the CTI shown in Fig~\ref{fig:cti} depend on the
 initial value of the CTI, we calculated the increase rate of the CTI
 ($\Delta$CTI) as a function of proton fluence at each incremental
 irradiation for various energies, shown in Fig~\ref{fig:delta_cti}.

 \subsection{Dependence of CTI on biased and unbiased devices}\label{sec:bias}

 On the satellite orbit including the ISS, high energy particles
 distribute far from uniform but concentrate in a very small area on the
 Earth. The most dense region of the high energy particles is so-called
 the South Atlantic Anomaly (SAA). During the passage of the SAA, the
 quality of data would be bad because of high background.  Therefore, if
 the performance degradation of CCDs depends on the biased (in
 operation) or unbiased (out of operation) condition, we could turn off
 CCDs during the passage of the SAA.

 We thus investigated the difference of device performance whether the
 device was biased or not during irradiation of 292\,keV protons.
 Figure~\ref{fig:cti_on-off} shows the $\Delta$CTI as a function of
 proton fluence. We found no significant difference between
 them. Therefore, the devices need not to be turned off during SAA only
 taking into account the degradation of CTI.

 \subsection{CTI for various devices and for various processes}\label{sec:wafer_result}

 As written in section~\ref{sec:wafer}, we fabricated four types devices
 to compare the difference of radiation hardness. 
 All devices were unbiased during the proton irradiation. The values of
 $\Delta$CTI obtained for these devices are shown in
 Fig~\ref{fig:cti_wafer}.  We found that $\Delta$CTI value obtained by
 epi-3 with notch is factor of $3-5$ times lower than that with epi-3
 without notch. Significant improvement is obtained although this value
 is slightly smaller than that of the geometrical ratio of notched area
 and other area. We thus decided to employ the notch structure for
 flight devices.

 There is no significant difference between epi-2 with notch and epi-3
 with notch, suggesting no differences in $\Delta$CTI for high and low
 resistivity wafers. We can therefore investigate the effect of the
 nitride oxide comparing between epi-3 without notch and epi-2 without
 notch but having the nitride oxide. There were very little differences
 between. Therefore, if the degradation of the dark current in the
 device having the nitride oxide is smaller than that without the
 nitride oxide, we will employ the nitride oxide for flight devices. The
 experimental results concerning about the dark current is described in
 the subsequent paper.

 \section{Discussion}

 \subsection{Proton Bragg curve}

 We found that protons having energies of 292 and 391\,keV seriously
 damaged HPK CCDs on the CTI performance. The degradation of the CTI
 obtained with protons having lower and higher energies is much less
 than those with 292 and 391\,keV protons. This strongly suggests that the
 low radiation-tolerant region inside the HPK CCD is located in
 relatively a narrow region.

 We calculated the Bragg curves of protons in Si. We employed {\sl
 Geant4} with the {\sl G4EMLOW0.3} data and considered the energy
 straggling due to the Al degrader of 5\,$\mu$m in thickness.
 Figure~\ref{fig:bragg_curve} ({\it upper}) shows the energy loss of
 protons as a function of depth of Si. The dotted line represents the
 minimum energy to displace Si atoms ($\simeq 6$ eV $\rm
 \AA^{-1}$)~\cite{sze}. The energy deposition due to 292 and 391\,keV
 protons are concentrated at the depth of $2-4\,\mu$m inside Si. In this
 depth, the energy deposition of protons with other energies is less
 than those of 292 and 391\,keV. Therefore, the radiation tolerance at
 depth of $2-4\,\mu$m is much lower than those in other region inside
 the HPK CCD.

 Figure~\ref{fig:bragg_curve} ({\it lower}) shows the schematic view of
 the cross section of the HPK CCD employed. Since the HPK CCD is a
 buried-channel type, the charge packet is transferred in a narrow
 region along the depth of the CCD. This transfer channel well coincides
 with the Bragg peak region. We thus conclude that the transfer channel
 of the CCD possesses the lowest radiation tolerance for protons. This
 result is consistent with the ACIS result but the serious proton energy
 is slightly different from our value. Prigozhin {\it et
 al.}~\cite{acis_damage} estimated the minimum proton energy to reach
 the buried channel to be somewhat higher than 50$-$70\,keV in order to
 penetrate the optical blocking filter, covering layer, and electrodes.
 Therefore, the thickness of the covering material is much thinner than
 our case, resulting that lower energy protons seriously affected the
 ACIS CCDs.

 As described in section~\ref{sec:wafer_result}, there is no difference
 in CTI values between CCDs fabricating from high resistivity wafer and
 those from low resistivity wafer. The acceptor doping concentration of
 our device is only an order of $10^{13}-10^{14}$\,cm$^{-3}$ and the
 difference between epi-2 wafer and epi-3 wafer is roughly an order of
 magnitude~\cite{ssc_em}.  Therefore, the probability that protons
 encounter Si atom is essentially the same between these devices.  Since
 the thickness of $n$-type layer is the same between them, their
 difference is the thickness of a depletion layer. It means that the
 location of the transfer channel is at the same depth between them.
 Our results are, therefore, expected if the radiation tolerance depends
 not on the depletion depth but on the transfer channel. This is
 consistent with the previous work~\cite{holland}.  We are now
 developing CCDs from newly-obtained epitaxial wafer having much higher
 resistivity than that of epi-3. Since, however, the location of the
 transfer channel of new CCDs is the same depth as current devices, we
 are convinced that we can apply these results to new CCDs.

 \subsection{Modeling the CTI degradation}

 As shown in Fig~\ref{fig:delta_cti}, the degradations of $\Delta$CTI
 are expressed as a linear function of the proton fluence. Since
 $\Delta$CTI is expressed as a linear function of the electron trap
 density~\cite{physics}, the formation of electron traps proportionally
 corresponds to proton fluence.  Values of $\Delta$CTI are fitted to a
 linear function of proton fluence. The best fit parameters, a slope and
 an intercept ($\Delta$CTI$_0$), are shown in Table~\ref{table:cti}.
 Figure~\ref{fig:slope} shows the slope obtained as a function of proton
 energy. Since the obtained values of slope correspond to an efficiency
 to create the electron trap, Fig~\ref{fig:slope} shows that 292\,keV
 protons most seriously affect the CTI degradation.

 As shown in Fig~\ref{fig:bragg_curve}, low energy protons deposit major
 part of their energy within a confined depth. The peak of the Bragg
 curve corresponds to the depth of 2.3\,$\mu$m in Si in the case of
 292\,keV protons. We thus assume there is a thin radiation-sensitive
 area within the CCD at depth of 2.3\,$\mu$m with thickness of
 0.05\,$\mu$m. We should note that 0.05\,$\mu$m is the shortest unit we
 can simulate. Ignoring the $\Delta$CTI degradation from other depths,
 we can calculate the energy deposition by protons that affect the CTI.
 Results are plotted in Fig~\ref{fig:slope} with filled circles
 normalized by value at 292\,keV.  For all proton energies, calculated
 values are much larger than those obtained.  As shown in
 Fig~\ref{fig:bragg_curve}, if the thickness of the radiation-sensitive
 region increases, the energy deposit of 391 or 522 keV protons becomes
 relatively larger than that of 292 keV protons. It drove the calculated
 values for 391 and 522 keV to be increased much more than current
 values, resulting the deviation from data to be more significant.  In
 this calculation, we assumed that the probability to create an electron
 trap is linearly proportional to the proton energy loss. This
 assumption leads to a large discrepancies between the data and the
 calculations.  Therefore, there may be some nonlinear effects in their
 probabilities.

 There are two types of process for proton energy loss: an ionization
 energy loss (IEL) and a nonionization energy loss (NIEL).  These two
 different forms of energy dissipation are translated into two major
 damage mechanisms for CCDs: an ionization damage and a bulk damage. The
 ionization damage leads to a flat-band shift which causes the operating
 voltage to be shifted. This damage is caused by all types of charged
 particles. On the other hand, energetic charged particles undergo
 Rutherford-scattering-type Coulombic interactions with the Si lattice
 structure. The energy deposited by the interacting ion is enough to
 pull a silicon atom out of its lattice position, forming an
 interstitial Si atom and a vacancy. The displaced atom, called the
 primary knock-on atom (PKA), may have sufficient energy to undergo
 collisions with lattice, producing more vacancies. NIEL is responsible
 for a part of the energy producing the initial vacancy-interstitial
 pairs and phonons.

 Ziegler {\it et al.}~\cite{iel} and Burke~\cite{niel} calculated the IEL and
 the NIEL, respectively. Based on their calculations, more than 98\,\%
 of incident proton energies ($E_p$ [keV]) release as the IEL for $E_p
 \ge 100$\,keV.  For a proton of relativistic energies, the NIEL is
 almost constant whereas with lower energies the NIEL has a 1/$E_p$
 dependence.  This suggests that the probability to create displacements
 is not linearly proportional to the total energy loss but is
 proportional to $E_p^{-\gamma}$.  We then fitted the function
 $E_p^{-\gamma}$ to the results of NIEL calculated by Burke.  We found
 that $\gamma$ can be approximated to be $\simeq 0.76$ at the energy
 range of $100 \,{\rm keV} \le E_p \le 4 $\,MeV.

 In order to take into account the nonlinear effect in creating traps
 due to the NIEL, we need to employ not the incident proton energy but
 the energy at the depth of 2.3\,$\mu$m. We calculated the energy
 reduction of $E_p$ during the passage of 2.3\,$\mu$m in Si with {\sl
 Geant4}.  We then calculated the fraction of the NIEL among the total
 energy loss with taking into account of the energy dependence of the
 NIEL for each reduced $E_p$.  We normalized the fraction of the NIEL
 for each proton energy by that of 292 keV and took them into account
 for the previous calculations. Results are shown by filled squares in
 Fig~\ref{fig:slope}. Our calculations considering the NIEL represent
 the data obtained. However, values of slope measured suddenly decreases
 as $E_p$ increases whereas they cannot be reproduced by our
 calculations. In our model, we only consider the NIEL which represent
 the energy deposition as the initial vacancy-interstitial pairs and
 phonons. If the energy of PKA is large enough to undergo collisions
 with Si atoms, the number of vacancies increase. Therefore, to take
 into account the spectrum of PKA and collisions between PKA and Si
 atoms is important for future modeling.

 Empirical relations between the slope of the $\Delta$CTI versus the
 proton energy are described as:

 \begin{eqnarray}\label{eqn:cti}
  {\rm slope}(E_p\,{\rm [keV]}) & = & 1.2\times 10^{-10}\times E_p - 2.0 \times 10^{-11}
   \ \ \ {\rm for} \ E_p \le 292\, {\rm keV} \\
  {\rm slope}(E_p\,{\rm [keV]}) & = & 1.2\times 10^{-9}\times\exp(-E_p/6.6\times 10^{-2})
   + 3.0 \times 10^{-13}
   \ \ \ {\rm for} \ E_p \ge 292\, {\rm keV}
 \end{eqnarray}

 \noindent
 Solid lines in Fig~\ref{fig:slope} represent above empirical relations.
 For a given proton spectrum in orbit, we can calculate the $\Delta$CTI
 value by summing contributions from all proton energies.

 \subsection{Estimate the CTI for the MAXI mission}

 We found that low energy protons with energies of $290-400$\,keV
 seriously damaged the spectroscopic performance of the MAXI CCDs. The
 degradation of CTI as a function of mission life for the MAXI based on
 our experiments has been estimated.  There is a slit at the top of the
 SSC camera with a size of 5$\times$0.3\,mm$^2$ and the slat collimators
 just above the CCDs~\cite{sakano}. The thickness of the slat collimator
 is $\sim 100\,\mu$m, which is aligned by $\sim 3\,$mm pitch, resulting
 the field of view of each CCD to be $\sim 1.5\,^\circ$ square. Within
 the field of view, no shield protects devices whereas the column
 density at other directions on the camera is $\sim 2.5$\,g cm$^{-2}$,
 suggesting the proton component passing through the camera to be
 negligibly small.  We thus calculate the proton flux coming through
 $1.5\,^\circ \times 1.5\,^\circ$ area.

 We employed the proton flux described in the literature~\cite{ssp}, in
 which the attitude of the ISS is 500\,km and solar activity is the
 maximum.  The proton flux at 500\,km is the largest among attitudes
 expected for the ISS~\cite{ssp} and we therefore use it for the worst
 case analysis.  The number of proton at the solar minimum is factor of
 $\sim 2$ larger than that at the solar maximum.  We thus increase the
 proton flux with a factor of 1.5 as the average
 value. Figure~\ref{fig:estimated_cti} shows the CTI estimated for the
 MAXI as a function of its mission life. The dotted line shows the
 acceptable limit for the MAXI mission. Since the mission life of the
 MAXI is two years, the degradation of the CTI is well below the
 acceptable limit even for the worst case analysis. We therefore confirm
 the high radiation torelance of MAXI CCDs.

 \acknowledgement

 This work is partly supported by the Grant-in-Aid for Scientific
 Research by the Ministry of Education, Culture, Sports, Science and
 Technology of Japan (13874032, 13440062).

 \begin{halftable}
  \caption{Energy of proton irradiated.}
  \label{table:proton_energy}
  \begin{halftabular}{@{\hspace{\tabcolsep}%
   \extracolsep{\fill}}ccc} \hline
   upstream of the degrader & \multicolumn{2}{c}{downstream of the
   degrader} \\\cline{1-1} \cline{2-3}
   proton energy [keV] & proton energy [keV] & width [keV] $^a$\\\hline
   570 & 171 & 13 \\
   650 & 292 & 12 \\
   720 & 391 & 11 \\
   820 & 522 & 11 \\
   2200 & 2061 & 10 \\
   4000 & 3911 & 9 \\
   \hline
  \end{halftabular} \\
  $^a$ Width is shown in unit of a standard deviation.\\
 \end{halftable}

 \begin{halftable}
  \caption{Result of linear fit for $\Delta$CTI as a function of proton
  energy.}
  \label{table:cti}
  \begin{halftabular}{@{\hspace{\tabcolsep}%
   \extracolsep{\fill}}ccc} \hline
   proton energy [keV] & slope & CTI$_0$\\\hline
   171 & (\error{1.63}{0.04})$\times 10^{-13}$ &
    (\error{5.3}{0.9})$\times 10^{-6}$ \\ 
   292 &  (\error{1.35}{0.05})$\times 10^{-11}$ &
   $< 8 \times 10^{-6}$ \\ 
   391 & (\error{3.34}{0.03})$\times 10^{-12}$ &
    (\error{6.3}{0.2})$\times 10^{-5}$ \\
   522 &(\error{1.10}{0.01})$\times 10^{-12}$ &
   (\error{6.70}{0.08})$\times 10^{-5}$ \\ 
   2061 &(\error{3.43}{0.06})$\times 10^{-13}$ &
   (\error{5.7}{0.9})$\times 10^{-6}$ \\ 
   3911 &(\error{2.26}{0.07})$\times 10^{-13}$ &
   $< 2 \times 10^{-6}$ \\ 
   \hline
  \end{halftabular}
 \end{halftable}

 \begin{halffigure}
  \caption{(a) Schematic view of the CCD.  (b) The cross section of a CCD
  pixel along the horizontal and the vertical direction
  is shown for the normal CCD. The horizontal cross section for
  (c) notch CCD and (d) nitride CCD is shown. The potential profile for
  electron is also shown in bottom of (c).}
  \label{fig:ccd_architecture}
 \end{halffigure}

 \begin{halffigure}
  \caption{Experimental setup of the proton irradiation test. }
  \label{fig:degrador-ccd}
 \end{halffigure}

 \begin{halffigure}
  \caption{Single-event spectra of $^{55}$Fe sources for device (a) before
  irradiated, irradiated to (b) $1.04\times 10^7$ and (c) $1.11 \times
  10^8$\,protons cm$^{-2}$ having energy of 292\,keV. All spectra were
  extracted from the same number of frames and taken at $-100\,^\circ$C.}
  \label{fig:specs}
 \end{halffigure}

 \begin{halffigure}
  \caption{Readout noise as a function of proton fluence for 292\,keV
  and 3.91 MeV protons.}
  \label{fig:readout_noise}
 \end{halffigure}

 \begin{halffigure}
  \caption{Pulse height of $^{55}$Fe events as a function of transfer
  number taken at $-100\,^\circ$C for device (a) before
  irradiated, irradiated to (b) $1.04\times 10^7$ and (c) $1.11 \times
  10^8$ protons cm$^{-2}$ having energy of 292\,keV.}
  \label{fig:pha_rawy}
 \end{halffigure}

 \begin{halffigure}
  \caption{CTI as a function of proton fluence for various proton
  energies. }
  \label{fig:cti}
 \end{halffigure}

 \begin{halffigure}
  \caption{$\Delta$CTI as a function of proton fluence for various proton
  energies.}
  \label{fig:delta_cti}
 \end{halffigure}

 \begin{halffigure}
  \caption{$\Delta$CTI as a function of proton fluence for the device biased
  and unbiased during the proton irradiation.}
  \label{fig:cti_on-off}
 \end{halffigure}

 \begin{halffigure}
  \caption{$\Delta$CTI as a function of proton fluence for devices
  fabricating from various different wafer and different processes.}
  \label{fig:cti_wafer}
 \end{halffigure}

 \begin{halffigure}
  \caption{Bragg curves for various proton energies ({\it upper}) and the
  schematic view of the cross section of HPK CCD employed ({\it
  lower}). The dotted line in the {\it upper} figure shows the
  minimum energy to displace Si atoms.} \label{fig:bragg_curve}
 \end{halffigure}

 \begin{halffigure}
  \caption{Slope of $\Delta$CTI as a function of proton energy. Model
  calculations with and without taking into account the nonlinear effect
  due to NIEL are also plotted by filled circles and filled squares,
  respectively. Solid lines represent the empirical relations.}
  \label{fig:slope}
 \end{halffigure}

 \begin{halffigure}
  \caption{CTI estimated for the MAXI CCDs. The dotted line indicates the
 requirement for the MAXI. }
  \label{fig:estimated_cti}
 \end{halffigure}

 \makefigurecaptions

 \clearpage

 \includegraphics[clip,scale=.7]{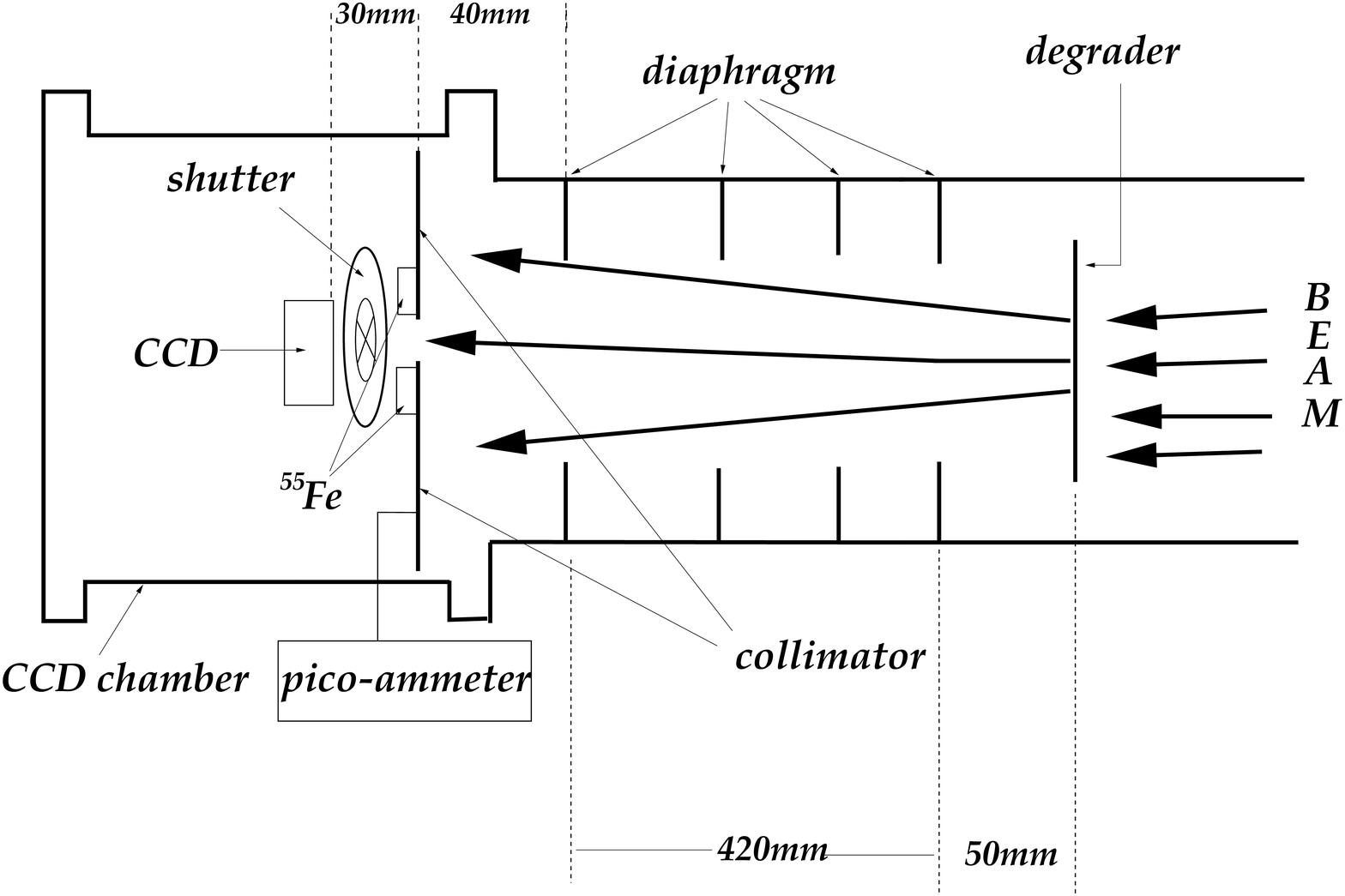}

 \clearpage

 \includegraphics[clip,scale=.4]{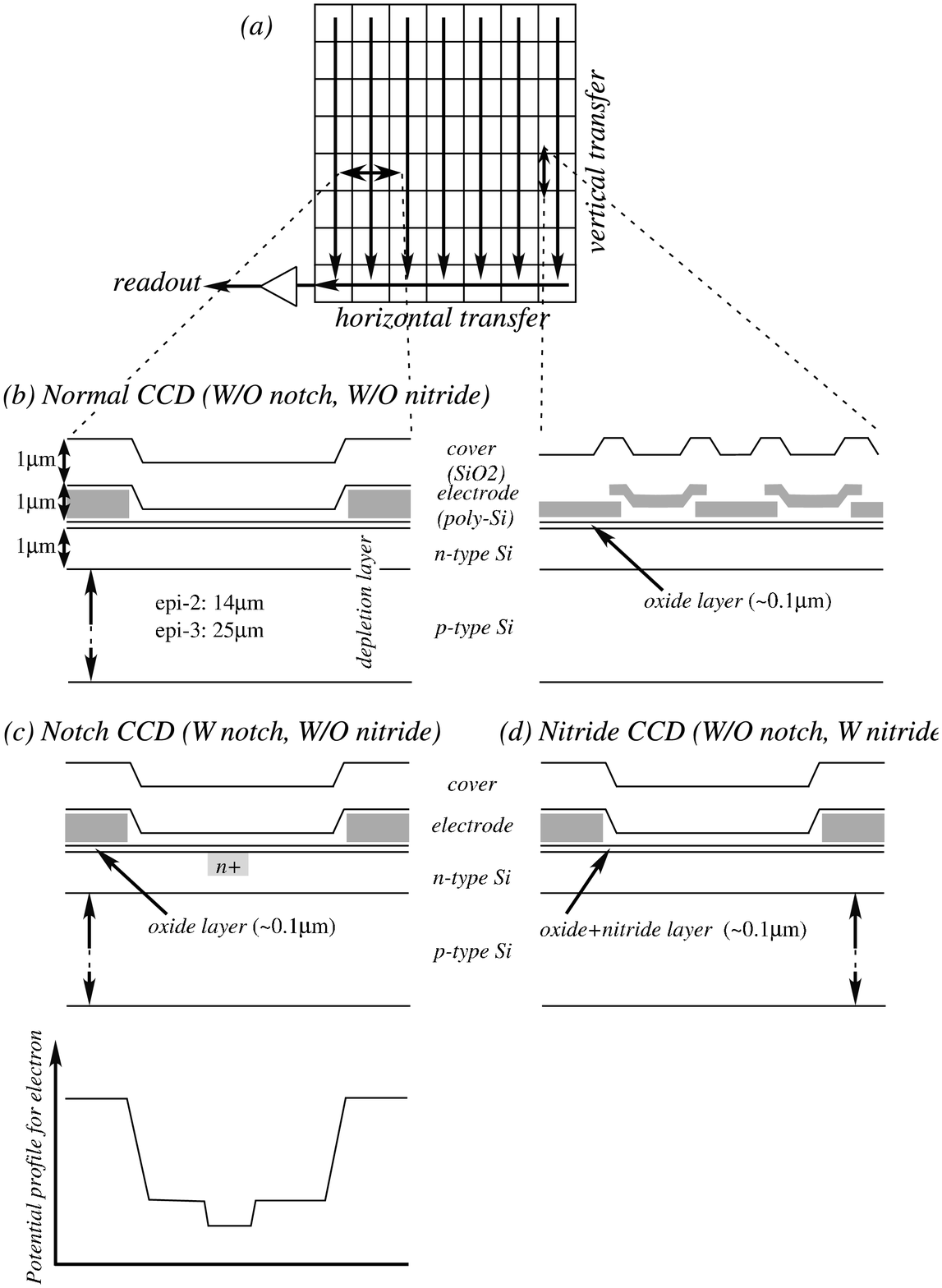}

 \clearpage

 \includegraphics[clip,scale=.4]{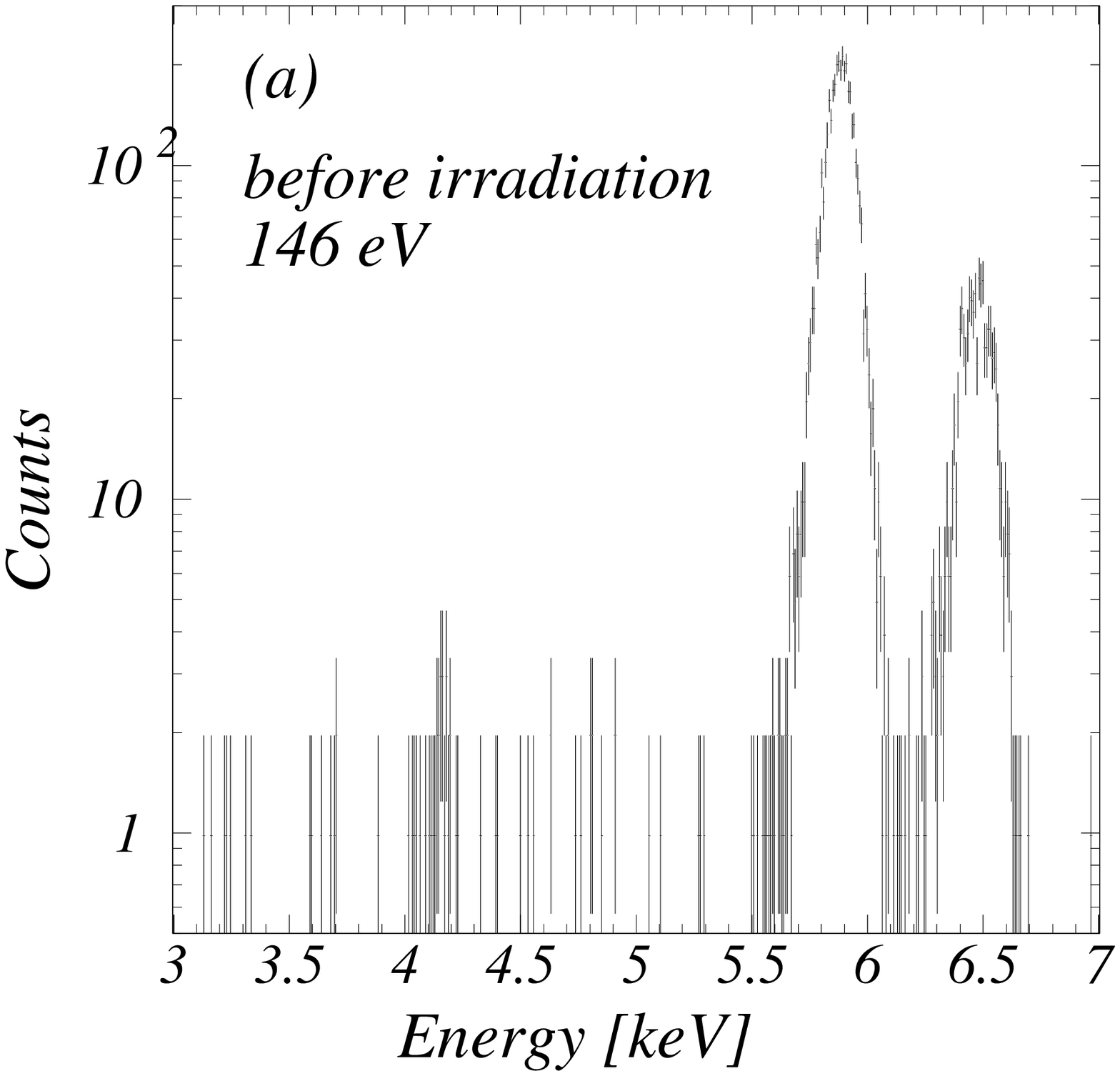}

 \includegraphics[clip,scale=.4]{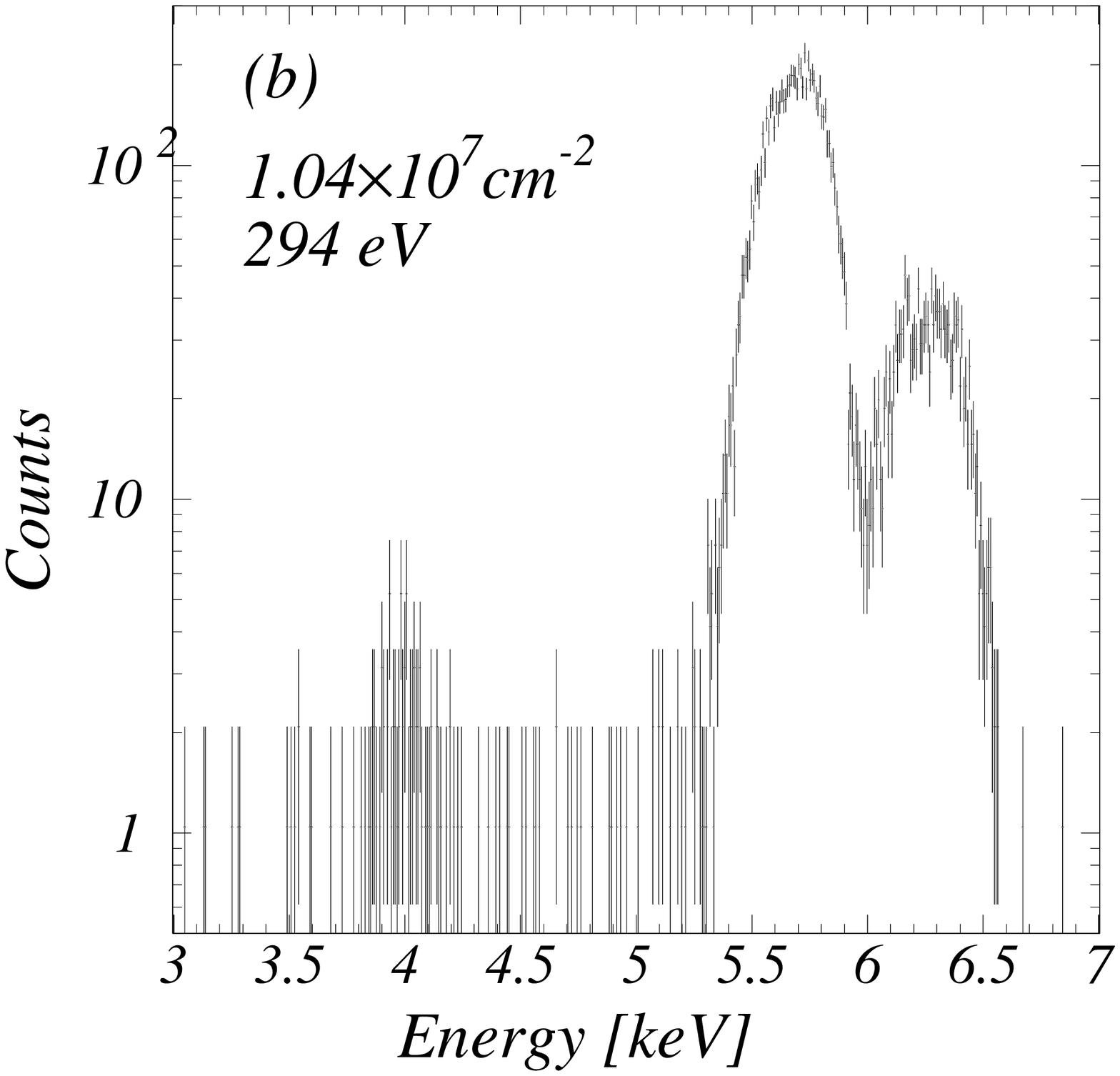}

 \includegraphics[clip,scale=.4]{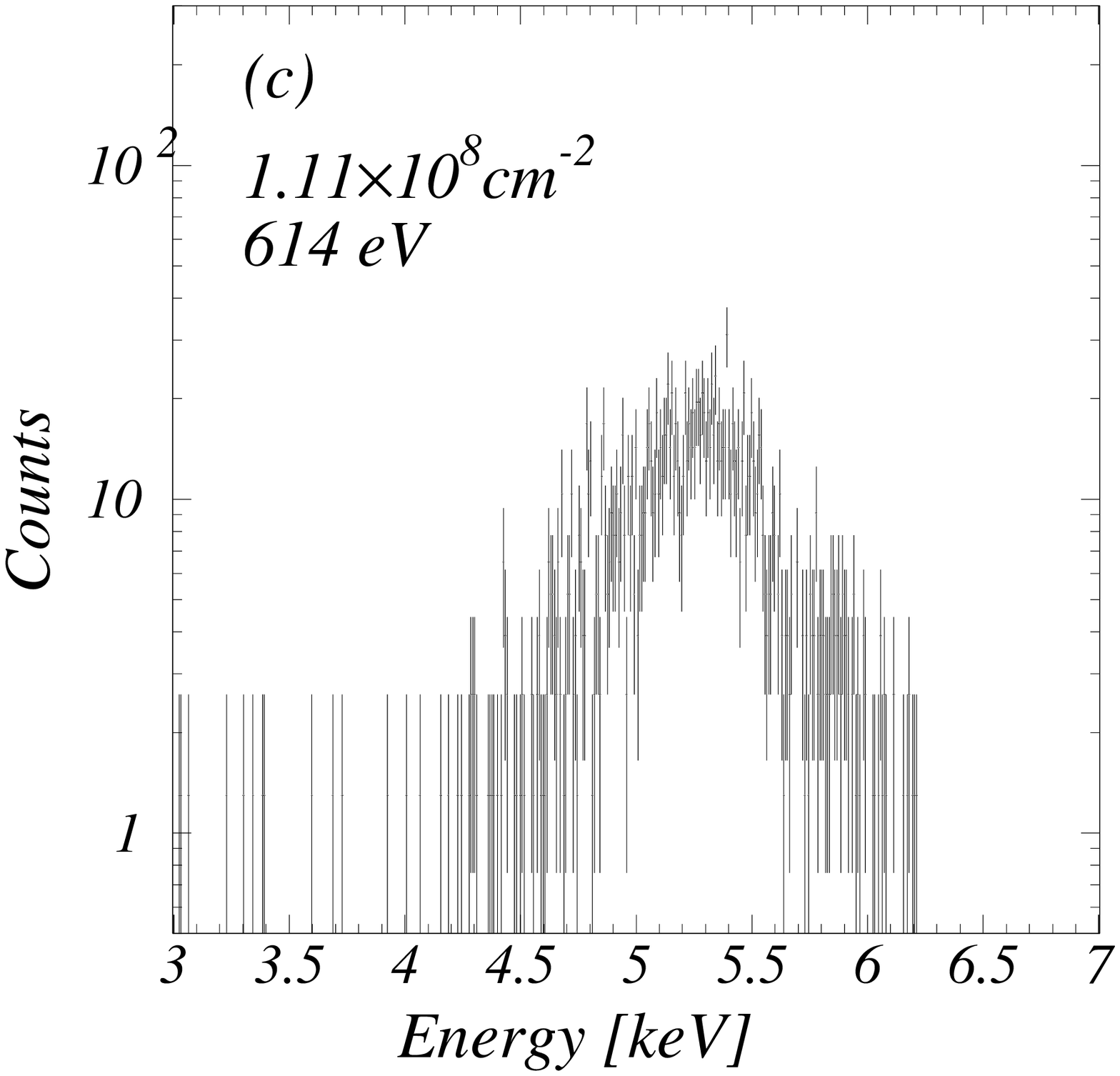}

 \clearpage

 \includegraphics[clip,scale=.8]{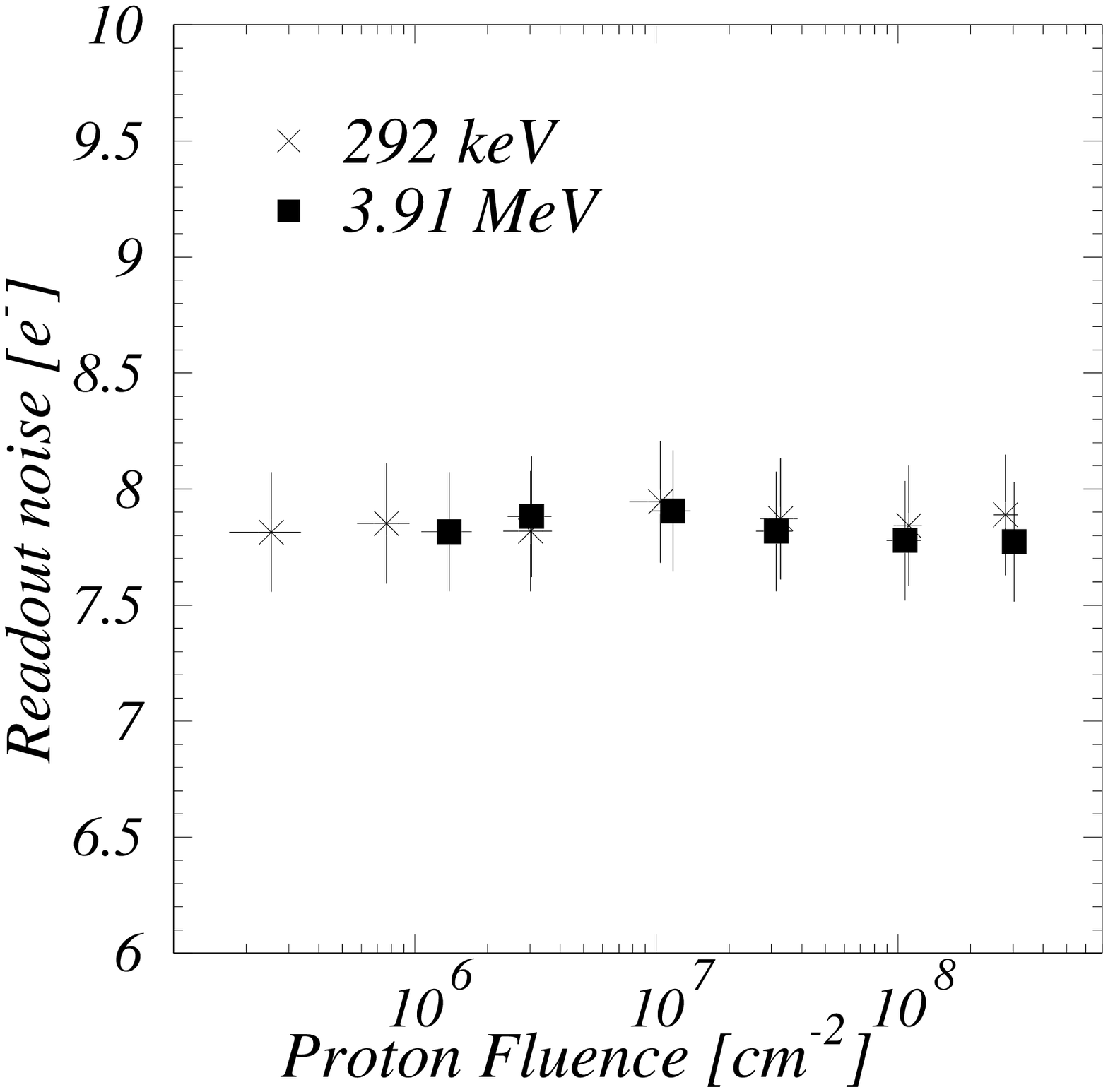}

 \clearpage

 \includegraphics[clip,scale=.4]{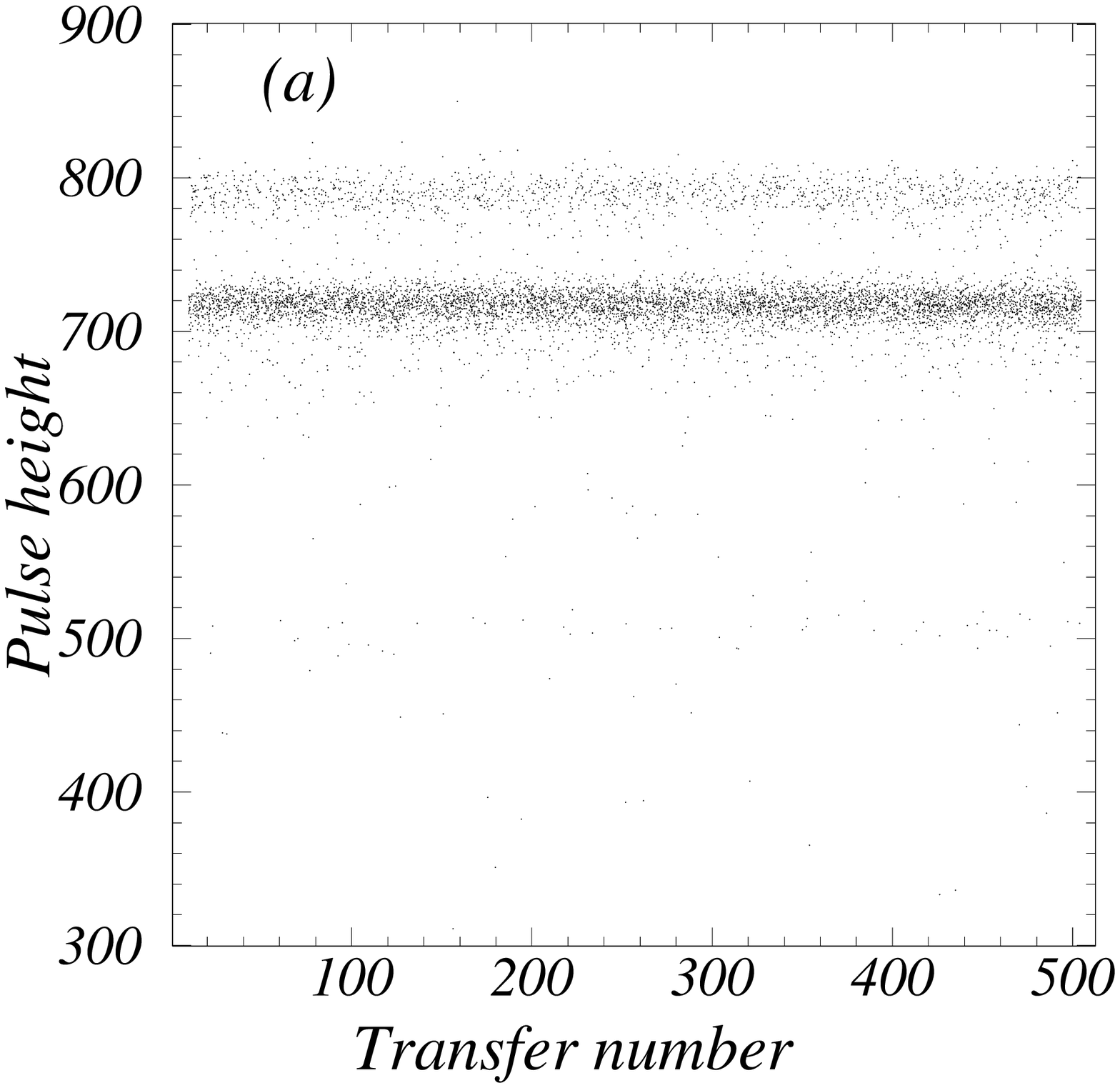}

 \includegraphics[clip,scale=.4]{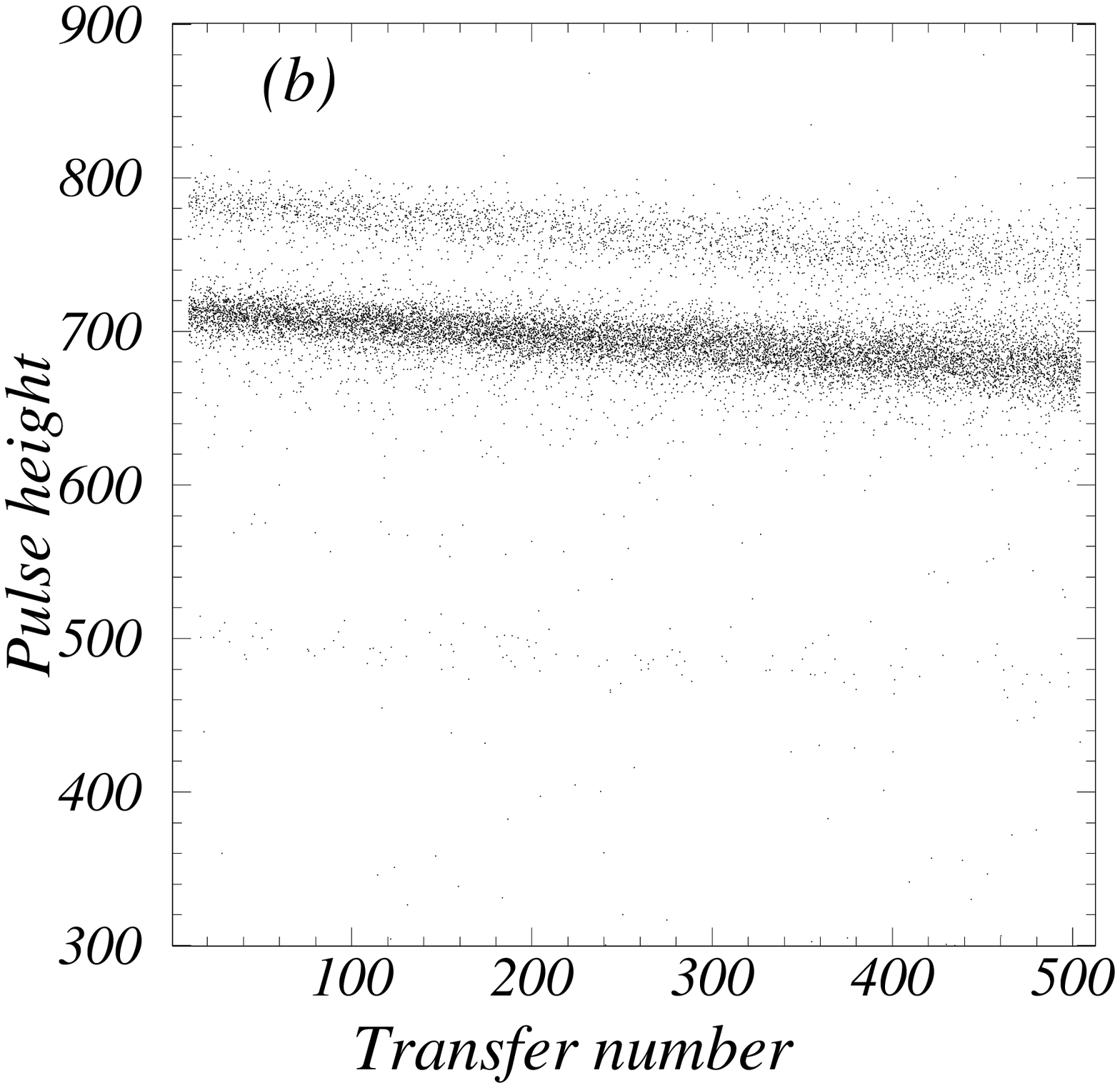}

 \includegraphics[clip,scale=.4]{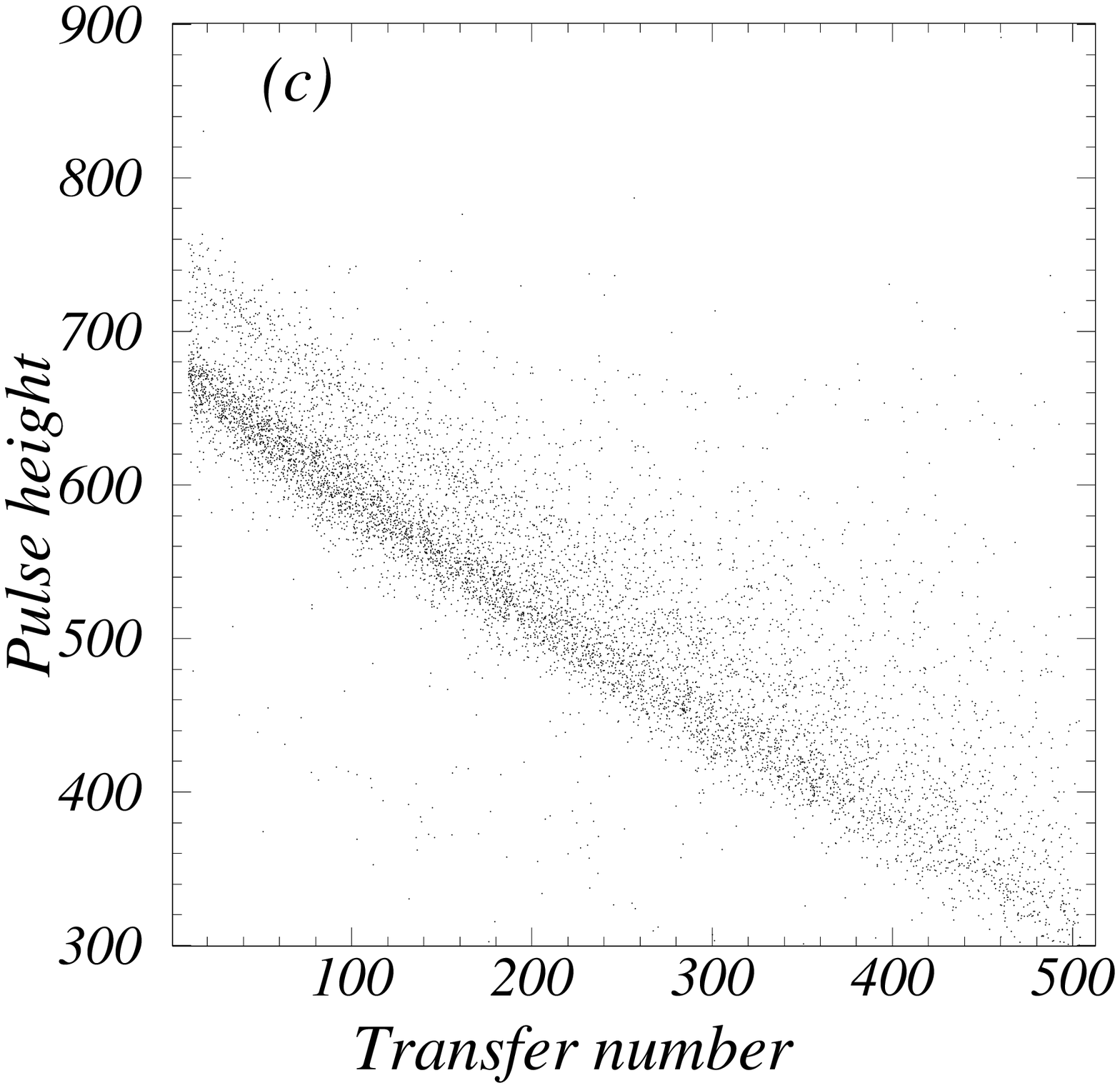}

 \clearpage

 \includegraphics[clip,scale=.8]{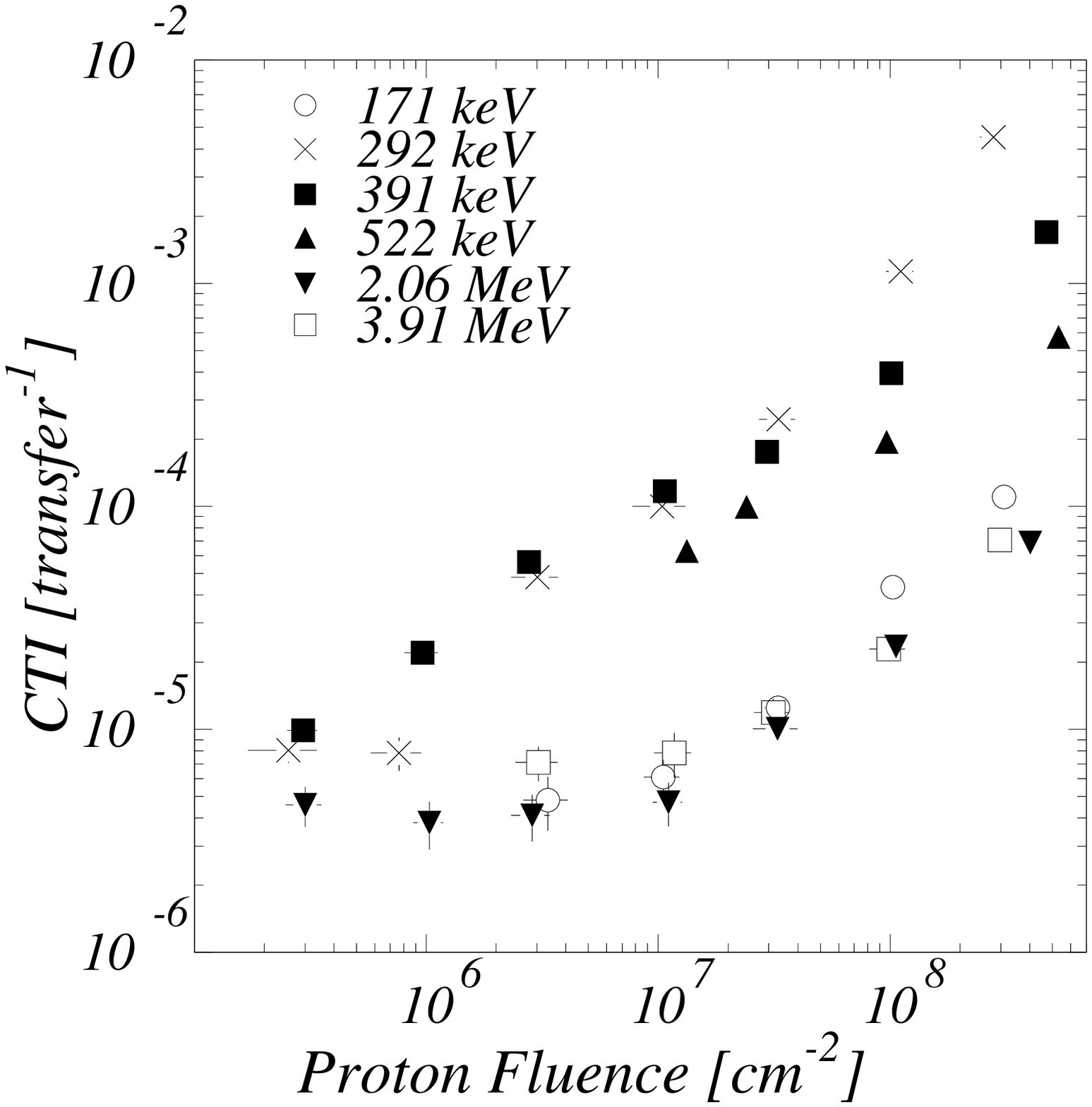}

 \clearpage

 \includegraphics[clip,scale=.8]{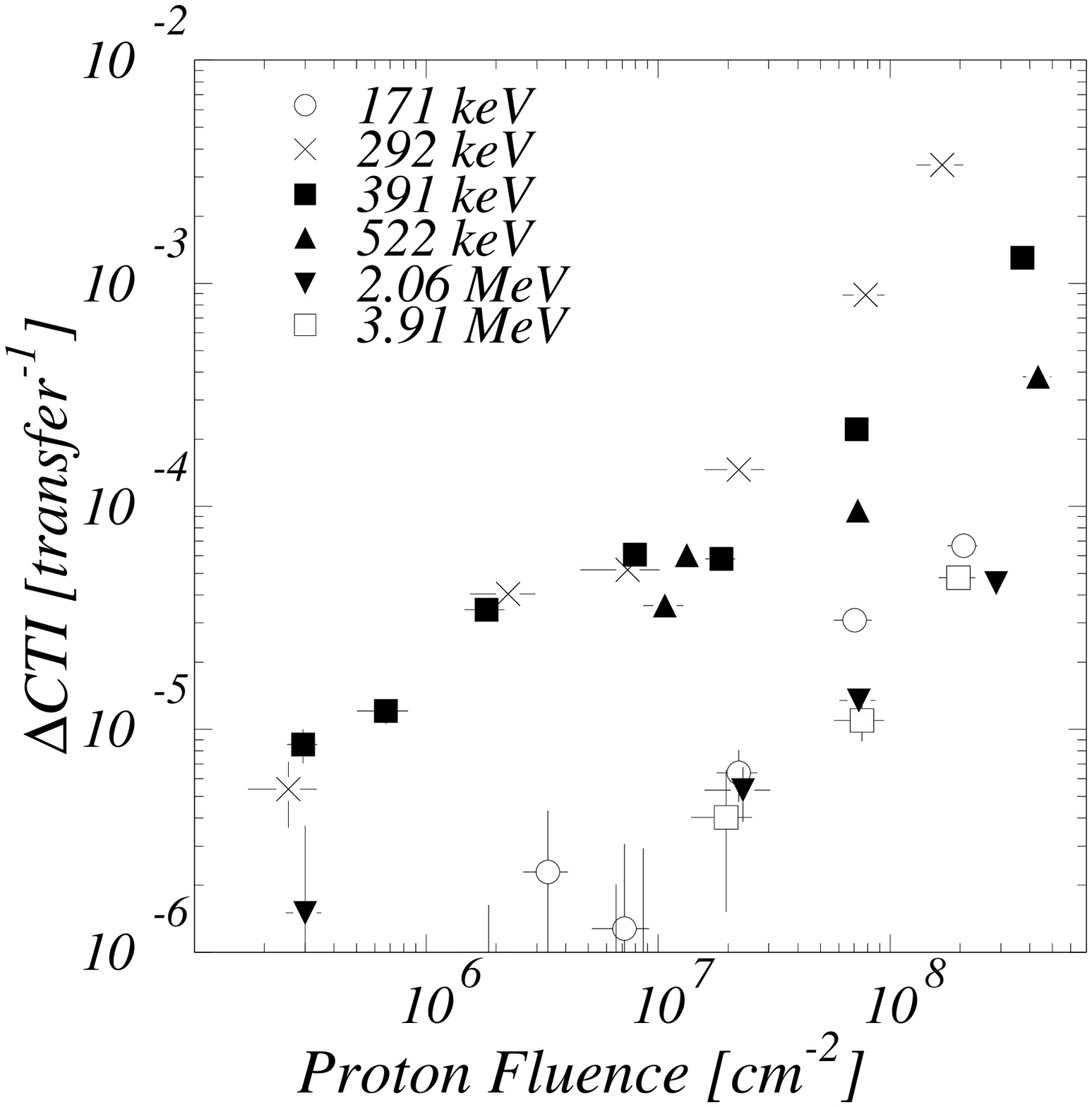}

 \clearpage

 \includegraphics[clip,scale=.8]{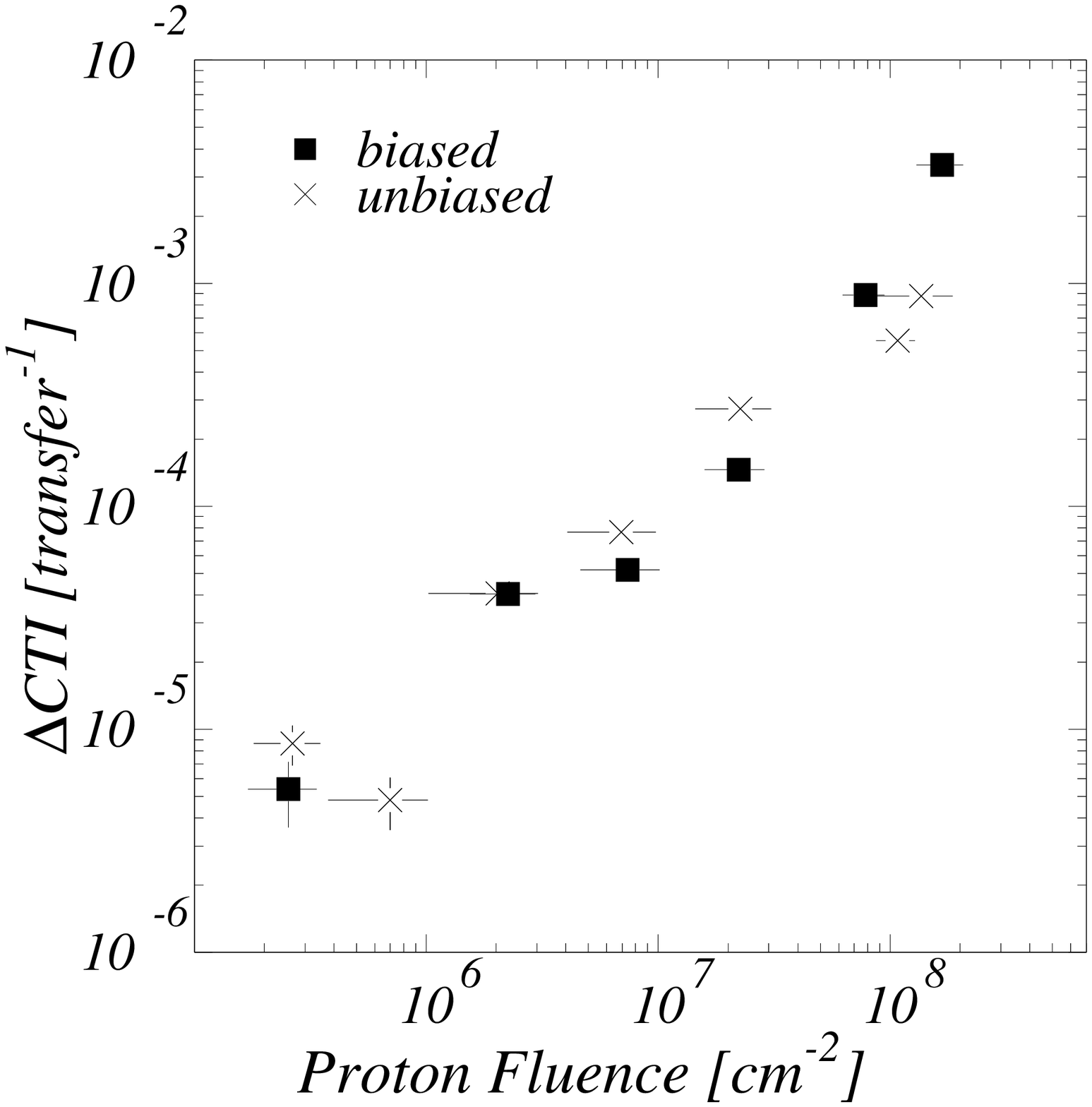}

 \clearpage

 \includegraphics[clip,scale=.8]{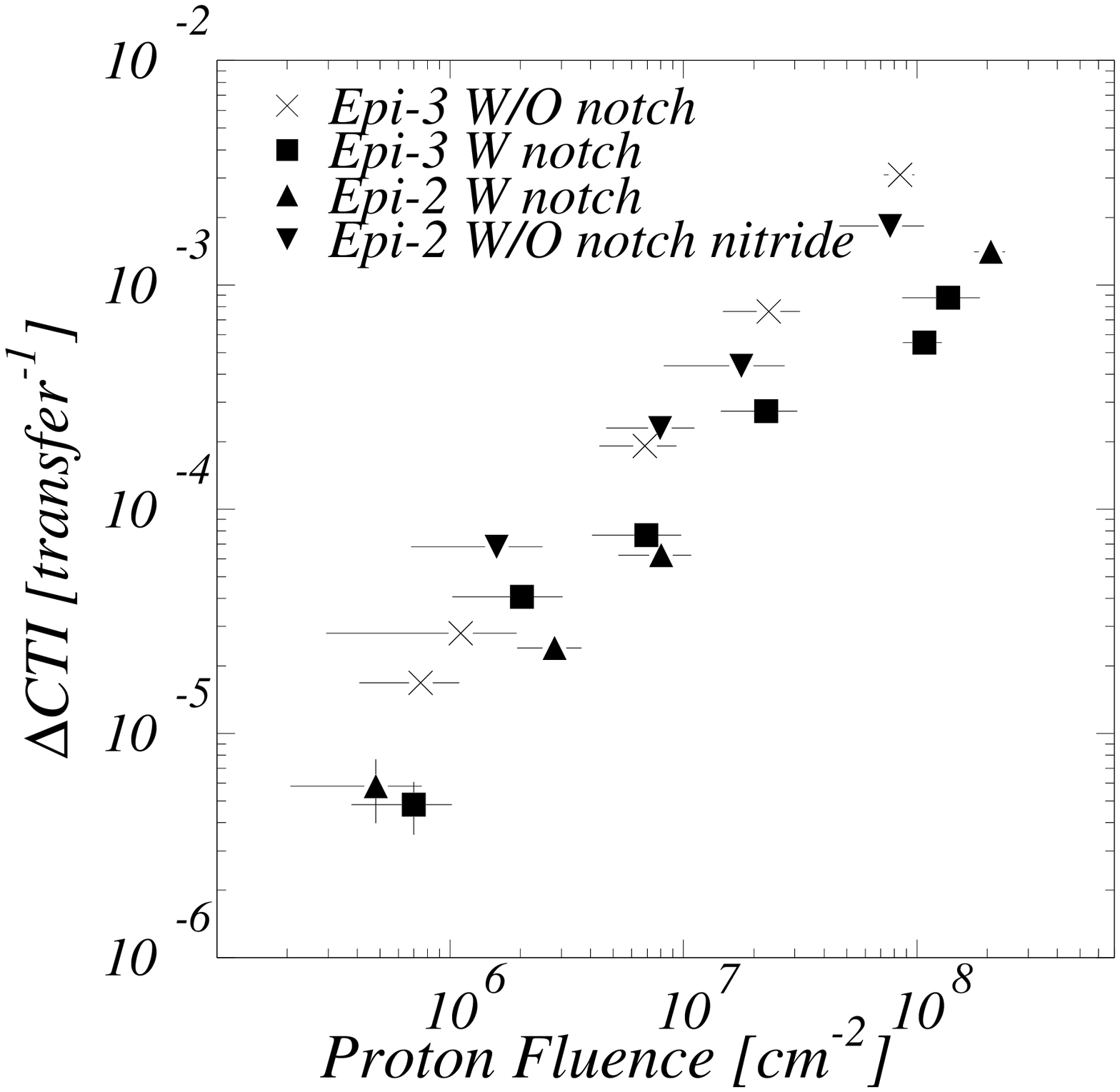}

 \clearpage

 \includegraphics[clip,scale=.4]{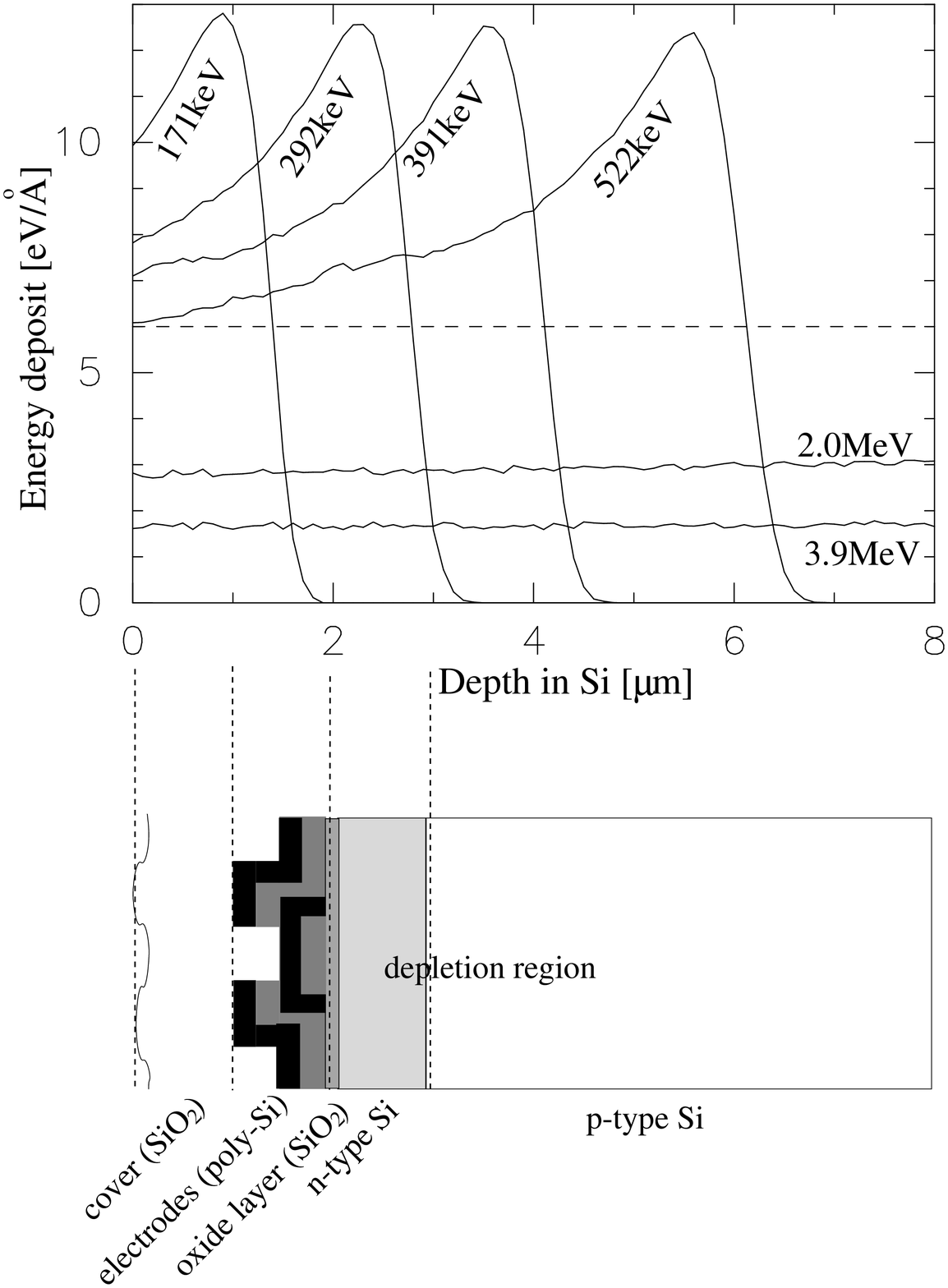}

 \clearpage

 \includegraphics[clip,scale=.8]{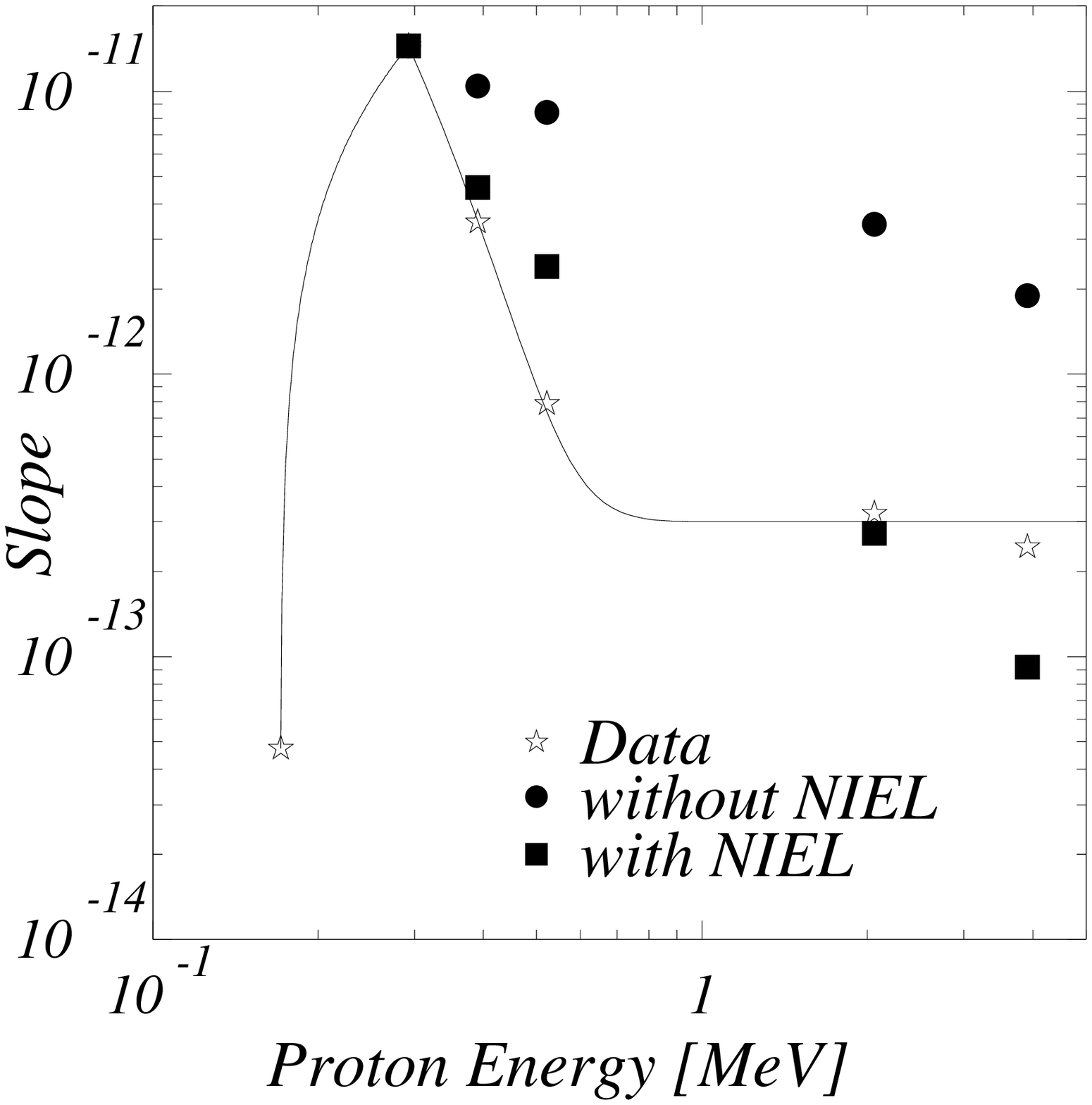}

 \clearpage

 \includegraphics[clip,scale=.5]{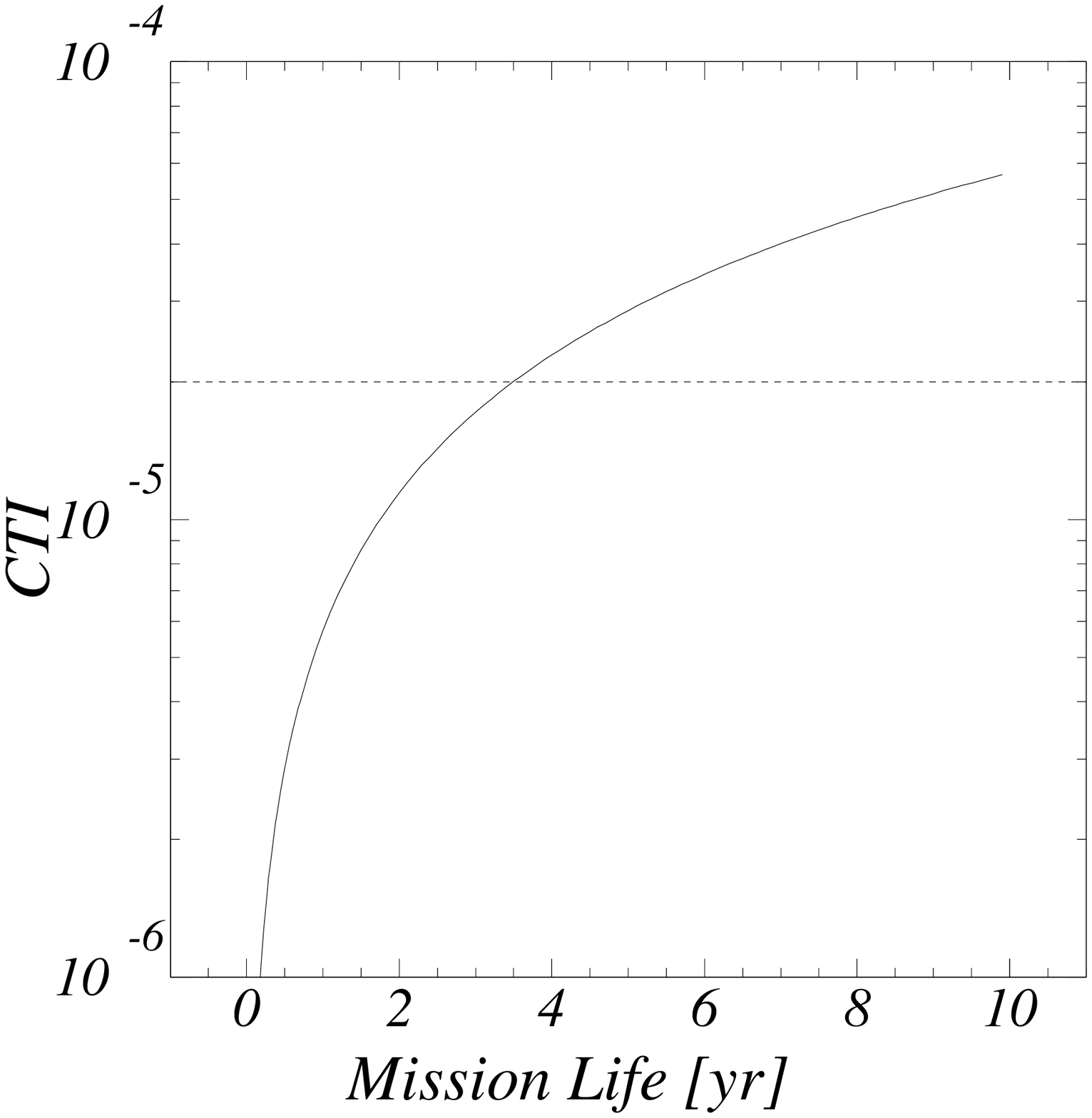}

 \end{document}